\documentclass[prl,aps,epsf,amsfonts,floats,twocolumn,amssymb,amsmath,groupedaddress,showpacs,floatfix,nofootinbib]{revtex4}

\usepackage{graphicx}
\usepackage[]{latexsym}
\usepackage{bm}

\newcommand{\be}{\begin{equation}}\newcommand{\ee}{\end{equation}}
\newcommand{\bea}{\begin{eqnarray}}\newcommand{\eea}{\end{eqnarray}}
\newcommand{\brr}{\begin{array}}\newcommand{\err}{\end{array}}
\newcommand{\bit}{\begin{itemize}}\newcommand{\eit}{\end{itemize}}
\newcommand{\ben}{\begin{enumerate}}\newcommand{\een}{\end{enumerate}}

\newcommand{\ba}{\begin{array}}
\newcommand{\ea}{\end{array}}

\def\lf{\left}

\def\non{\nonumber}\def\ran{\rangle}

\def\ri{\right}

\def\te{\theta}

\def\1{{_{1}}}\def\2{{_{2}}}

\def\noHe0{:\;\!\!\;\!\!:H_e(0):\;\!\!\;\!\!:}
\def\noHm0{:\;\!\!\;\!\!:H_\mu(0):\;\!\!\;\!\!:}

\def\lf{\left}

\def\non{\nonumber}
\def\ran{\rangle}

\def\ri{\right}

\def\te{\theta}

\def\1{{_{1}}}\def\2{{_{2}}}

\begin{document}

\title{Quantum vacuum,  dark matter, dark energy  and spontaneous supersymmetry breaking}

\author{ Antonio Capolupo\footnote{e-mail address: capolupo@sa.infn.it}}

 \affiliation{  Dipartimento di Fisica E.R.Caianiello and INFN gruppo collegato di Salerno,
  Universit\'a di Salerno, Fisciano (SA) - 84084, Italy}

\pacs{ }

\begin{abstract}

We study the vacuum condensate characterizing many physical phenomena. We  show that such a condensate may leads to non-trivial components of the dark  energy and of the dark matter and may induces the spontaneous supersymmetry breaking, in a supersymmetric context.  In particular, we consider the condensate induced by thermal states,  fields in curved space-time and  mixed particles.

\end{abstract}

\maketitle

\section{I. Introduction}

According to recent experimental data, the universe consists of $68\%$ of dark energy \cite{CMBR}-\cite{SNeIa} and  about of $27\%$ of dark matter. The rest is ordinary matter.
The dark energy and the dark matter problem have been analyzed in different ways \cite{odintsovreport}- \cite{kam6}, however, their solution  represents still a very big challenge.

Another object of study which has had a huge impact on contemporary physics is the Supersymmetry (SUSY) \cite{Fayet:1974jb}--\cite{Buchholz:1997mf}.
SUSY is a  symmetry of nature that  relates any boson to a fermion (called superpartner) with the same mass and internal quantum numbers, and vice-versa. However,   there is no  evidence for the existence of  the superpartners. Therefore, SUSY must be a broken symmetry, allowing for superparticles to be heavier than the corresponding Standard Model particles, or it must be ruled out as a fundamental symmetry.
Intensive study has been devoted to the analysis of the possibility of SUSY breaking.

 Here we report on recent results \cite{CapolupoDark} according to which  the  vacuum condensate characterizing many phenomena \cite{Hawking:1974sw}--\cite{Capolupo:2004pt}, such as Hawking effect and fields in curved space, can explain the origin of  dark energy and matter components   \cite{CapolupoDark} and can describe the spontaneous SUSY breaking \cite{Capolupo:2010ek}--\cite{Capolupo:2012vf2}.
In fact, all the  phenomena inducing condensates
 have  a non-zero vacuum energy which cannot be removed by use of the normal ordering procedure.
 The  origin of the  non trivial vacuum energy  is due to the fact that the physical vacuum of such systems is a condensate of couples of particles and antiparticles which generate a positive value of the zero point energy.
Such an energy can contribute to the dark sector of the universe and, in a supersymmetric context, induces  the spontaneous SUSY breaking  \cite{Capolupo:2010ek}.

In particular, we analyze   thermal states, fields in curved space and mixed particles and we show
 that dark matter components can be originated
by the thermal vacuum of the hot plasma present at the center of a galaxy cluster (intracluster medium),
by  vacuum fluctuations of fields in curved space  \cite{maroto} and by
the flavor neutrino vacuum \cite{CapolupoDark}.
Moreover, we show that dark energy contributes are given by  vacuum condensates induced by the axion-photon mixing and by superpartners of mixed neutrinos  \cite{CapolupoDark}.

We then consider the free Wees-Zumino model as a supersymmetric field theory, and we show that the presence of nonvanishing vacuum energy at the Lagrangian level implies that SUSY is spontaneously broken by the condensates.
Next experiments using atomic systems  characterized by vacuum condensate, could  test our conjecture.

In Sec.II, we introduce  the Bogoliubov transformations in QFT. In Sec.III, we compute the energy density and pressure of vacuum condensates induced by a generic Bogoliubov transformation for boson and fermion fields. In Secs.IV, V and VI, we present  the contribution given to the energy of the universe by thermal states, with reference to the Hawking and Unruh effects, by fields in curved space and by  particle mixing phenomena, respectively. The SUSY breaking induced by vacuum condensate is presented in  Sec.VII and  Sec.VIII is devoted to the  conclusions.

\section{II. Bogoliubov transformation and vacuum condensate}

In the context of   QFT, the  Bogoliubov transformations  \cite{Umezawa:1993yq} describe  phenomena such as the Hawking-Unruh effect \cite{Hawking:1974sw,Unruh:1976db}, the Schwinger effect \cite{Schwinger:1951nm}, the BCS theory of superconductivity \cite{Bardeen:1957mv},  the Thermo Field Dynamics \cite{Takahasi:1974zn}--\cite{Umezawa:1993yq}, the QFT in curved spacetimes \cite{Birrell} and the particle mixing phenomena \cite{Blasone:2002jv}-\cite{Capolupo:2004pt},
\cite{Capolupo:2006et}-\cite{Capolupo:2007hy3}.

A Bogoliubov transformation  for bosons (similar discussion hold for fermions) assumes the form
\bea\label{Bog12}\non
 {  a}_{\mathbf{k}}( \xi, t) &=& U^{B}_{\mathbf{k}} \, a_{\mathbf{k}}(t) - V^{B}_{-\mathbf{k}}  \,a^{\dagger}_{-\mathbf{k}}(t)\,,
\\
 {  a^{\dag}}_{-\mathbf{k}}( \xi,  t) &=& U^{B *}_{-\mathbf{k}} \, a^{\dag}_{-\mathbf{k}}(t) - V^{B *}_{\mathbf{k}}  \,a_{\mathbf{k}}(t)\,,
\eea
with $a_{\mathbf{k}}(t) = a_{\mathbf{k}} e^{-i\omega_{k}t}$,
annihilators, such that $a_{\mathbf{k}}|0\rangle_{B} =   0$ and $\omega_{\mathbf{k}}=\sqrt{k^2 + m^2}$.
The  coefficients satisfy the conditions
$
U^{B}_{\mathbf{k}} = U^{B}_{-\mathbf{k}}$, $ V^{B}_{\mathbf{k}} = V^{B}_{-\mathbf{k}}$, $|U^{B}_{\mathbf{k}}|^2 - |V^{B}_{\mathbf{k}}|^2 = 1\,,
$ and similar for fermions.
The parameter $ \xi$ depends on the system one considers.

The  transformations (\ref{Bog12}) can be rewritten in terms of a generator $J(\xi,   t)$ as
$
 {a} _{\mathbf{k}}( \xi,  t) = J ^{-1} (\xi, t)\,a_{\mathbf{k}}(t) J (\xi,  t)\,,
$
where $J (\xi,  t)$ has the property, $J ^{-1}(\xi)=J (-\xi)$.
The vacua annihilated by ${a} _{\mathbf{k}}( \xi,  t)$, denoted with $ |0( \xi,  t)\rangle $, are related to the original vacua  $|0\rangle$  by
$ |0( \xi,  t)\rangle = J^{-1} (\xi,  t)|0\rangle  $.
 Such a relation is a unitary operation in quantum mechanics, where $\mathbf{k}$ assumes a discrete range of values and there is a finite or countable number of canonical commutation relations. On the contrary, in QFT, $\mathbf{k}$ assumes a continuous infinity of values and  the relation
$ |0( \xi,  t)\rangle = J^{-1} (\xi,  t)|0\rangle  $ is not a unitary transformation any more.
In this case,
 $| 0 (\xi,t) \rangle_{B}$ and  $|0\rangle_{B}$ are unitarily inequivalent   and  the physical vacua of the  systems described by Bogoliubov transformations \cite{Hawking:1974sw}--\cite{Birrell},  are the $|0(\xi,t) \rangle $ ones \cite{Umezawa}--\cite{Umezawa:1993yq}.
Notice that
  $|0(\xi,t)\rangle $  has a condensate structure, i.e.
\bea\label{cond1}
 && \langle 0(\xi,t)| a^{\dagger}_{\mathbf{k}} a_{\mathbf{k}} |0(\xi,t)\rangle    = |V_{\mathbf{k}} |^2 \,\,\,\,\,\,\, \label{cond2},
\eea
which induces an energy momentum tensor different from zero for $|0(\xi,t)\rangle_{\lambda} $.

\section{III. Energy-momentum tensor of vacuum condensate}

In order to derive the state equation of the vacuum condensates $| 0 (\xi,  t)\rangle_{\lambda}$,  ($\lambda =B,F$) and their contributions to the energy density,
one computes the expectation value of  the free energy momentum tensor densities $T^{\mu\nu}(x)$ for real scalar fields
  and for Majorana fields  on
$| 0 (\xi,  t)\rangle_{\lambda}$,
\bea
\Xi^{\lambda}_{\mu \nu}(x) & \equiv & _{\lambda}\langle 0 ( \xi, t)|: T^{\lambda}_{\mu \nu}(x): | 0 (\xi,  t)\rangle_{\lambda}
\\\non
& = & _{\lambda}\langle 0 ( \xi, t)| T^{\alpha}_{\mu \nu}(x) | 0 (\xi,  t)\rangle_{\lambda}
- _{\lambda}\langle 0  | T^{\lambda}_{\mu \nu}(x) | 0  \rangle_{\lambda}\,,
\eea
where, $:...:$,  denotes the normal ordering with respect to the original vacuum $|0 \rangle_{\lambda}$.
Since the off-diagonal components of $\Xi^{\lambda}_{\mu \nu}(x)$ are zero, $\Xi^{\lambda}_{i,j}(x) = 0$, for $i \neq j$,
  the condensates   behave as a perfect fluid and one can define  their  energy density and pressure as \cite{CapolupoDark}
\bea
\rho^{\lambda} & = &  \langle 0 ( \xi, t)|: T_{0 0}^{\lambda}(x): | 0 (\xi,  t)\rangle\,,
\\
p^{\lambda} & = & \langle 0 ( \xi, t)|: T_{j j}^{\lambda}(x): | 0 (\xi,  t)\rangle\,,
\eea
respectively.

For bosons, one has \cite{CapolupoDark}
\begin{widetext}

\bea \label{T00Bos}
\rho_{B} & = &  \frac{1}{2}  \langle 0 ( \xi, t)|   : \Big[\pi^{2}( x) + \lf(\vec{\nabla} \phi( x)  \ri)^{2}
+ m^{2} \phi^{2}( x) \Big]: |0( \xi, t) \rangle\,;
\\ \label{TjjBos}
p_{B} & = &    \langle 0 ( \xi, t)| :\Big(\lf[ \partial_{j} \phi( x) \ri]^{2}+ \frac{1}{2}\Big[\pi^{2}( x)
 -   \lf(\vec{\nabla} \phi( x)  \ri)^{2}
 - m^{2}
\phi^{2}( x)  \Big] \Big): |0 ( \xi, t) \rangle\,,
 \eea

\end{widetext}
 and, in the particular case of the isotropy of the momenta, $k_1 = k_2 =k_3$,
 such that,
  $ \lf[ \partial_{j} \phi( x) \ri]^{2} = \frac{1}{3} \lf[\vec{\nabla} \phi( x)  \ri]^{2}$,
the energy density, the pressure and the state equation  are
\bea\label{energy-Bos}
\rho_{B}   & = & \frac{1}{2 \pi^{2}}  \int_{0}^{\infty} dk k^{2}\omega_{k } |V_{ k}^{B}|^{2}\,,
\\\non
\\\label{pressure-Bos}\non
p_{B}  & = & \frac{1}{6 \pi^{2}}   \int_{0}^{\infty} dk k^{2}\, \Big[  \frac{k^2}{\omega_k}
|V_{ k}^{B}|^{2}
\\
& - &  \lf( \frac{  k^2}{ \omega_k} + \frac{3 m^2}{2 \omega_k} \ri) |U_{ k}^{B}||V_{ k}^{B}| \cos (\omega_{k} t)  \Big] \,,
\\\non
w_{B} & = & \frac{1}{3}\frac{\int d^{3}\, \mathbf{k} \frac{k^2}{\omega_k}|V_{k}^{B}|^2 }{\int d^{3} \mathbf{k}\, \omega_k |V_{k}^{B}|^2}
\\
& - &   \frac{1}{3}\frac{\int d^{3} \mathbf{k}\, \lf( \frac{  k^2}{ \omega_k} + \frac{3 m^2}{2 \omega_k} \ri)  U_{k}^{B} V_{k}^{B} \cos (\omega_k t) }{\int d^{3} \mathbf{k}\, \omega_k |V_{k}^{B}|^2}\,,
\eea
respectively.

For fermions, the  energy density and   the pressure   are \cite{CapolupoDark}

\bea \label{T00Ferm}\non
\rho_{F} & = &  \frac{1}{2}   \langle 0 ( \xi, t)|   : \Big[-i \bar{\psi}\, \gamma_{j} \partial^{j}\, \psi + m \bar{\psi} \psi \Big]: |0( \xi, t) \rangle\,;
\\
\\ \label{TjjFerm}
p_{F} & = &     \langle 0 (\xi, t)| :\Big(  \frac{i}{2} \bar{\psi} \,\gamma_{j} \overleftrightarrow{\partial_{j}} \psi  \Big): |0 (\xi, t) \rangle\,.
 \eea
and explicitly one has

\bea\label{energy-Ferm}
\rho_{F}  &= & \frac{1}{\pi^{2}} \int_{0}^{\infty} dk k^{2} \omega_{k }  |V_{ k}^{F}|^{2}\,,
 \\\label{pressure-Ferm}
 p_{F}    &= & \frac{1 }{3 \pi^{2}}    \int_{0}^{\infty} dk \frac{k^{4}}{\omega_{k }}|V_{ k}^{F}|^{2}\,,
 \\
w_{F} & = & \frac{1}{3}\frac{\int d^{3}\, \mathbf{k} \frac{k^2}{\omega_k}|V_{k}^{F}|^2 }{\int d^{3} \mathbf{k}\, \omega_k |V_{k}^{F}|^2}\,.
\eea

We also note that, being $J^{-1}(\xi,t)  =  J^{\dag}(\xi,t) = J(-\xi,t)$, one can write
\bea\label{tensor}\non
&&_{\lambda}\langle 0 ( \xi, t)|: T^{\lambda}_{\mu \nu}(x): | 0 (\xi,  t)\rangle_{\lambda} =
\\&& = _{\lambda}\langle 0  |J^{-1}_{\lambda}(-\xi,t) : T^{\lambda}_{\mu \nu}(x): J_{\lambda}(-\xi,t)| 0  \rangle_{\lambda} \,.
\eea

In the following,  we denote with $\Theta(-\xi,x) = J^{-1}(-\xi,t) \Theta(x) J(-\xi,t) $ the operators  transformed by   $J (-\xi,t)$.
 All the equations above presented hold for many systems. The explicit form of the Bogoliubov coefficients
specifies the particular system.

\section{IV. Vacuum contributions of thermal states, Hawking and Unruh effects}

In the context of the Thermo Field Dynamics (TFD) \cite{Takahasi:1974zn}--\cite{Umezawa:1993yq},
  the physical vacuum of systems at non-zero temperature is the thermal vacuum state $|0(\xi(\beta))\rangle_{\lambda}$, where $\beta \equiv 1/(k_{B}T)$, $k_{B}$ is the Boltzmann constant and $\lambda = B,F$. The state $|0(\xi(\beta))\rangle_{\lambda}$ is obtained by means of a Bogoliubov transformation similar to the ones presented in Section I, (for details see
\cite{Umezawa:1993yq}) and
   the thermal statistical average ${\cal N}_{\chi_{\bf k}}(\xi)$ is given by
  ${\cal N}_{\chi_{\bf k}}(\xi) \,=\, _{\lambda}\langle 0(\xi(\beta))| N_{\chi_{\bf k}} |0(\xi(\beta))\rangle_{\lambda}$, where $N_{\chi_{\bf k}} = \chi^{\dag}_{\bf k} \chi_{\bf k} $, ($\chi = a$ for bosons and $\alpha$ for fermions) is the number operator \cite{Umezawa}.

The thermal Bogoliubov coefficients are given by
$U^{T}_{{\bf k}} = \sqrt{\frac{e^{\beta \omega_{\bf k} }}{e^{\beta \omega_{\bf k } }\pm 1}}$ and
$V^{T}_{{\bf k}} = \sqrt{\frac{1}{e^{\beta \omega_{\bf k} } \pm 1}}$, with $-$ for bosons and $+$ for fermions,
and $\omega_{\bf k } = \sqrt{k^{2} + m^{2} }$.
Such coefficients, used in Eqs.(\ref{energy-Bos}), (\ref{pressure-Bos}) and  (\ref{energy-Ferm}), (\ref{pressure-Ferm}) give the contributions of the thermal vacuum states to the energy  and pressure.
In particular, for temperatures of order of  the  cosmic microwave radiation, i.e. $T = 2.72 K$, one find that photons and particles with masses of order of $(10^{-3} - 10^{-4})eV$  contribute   to the energy radiation with $\rho \sim 10^{-51}GeV^{4} $ and state equations, $w = 1/3$ \cite{CapolupoCMB:2016}. On the other hand,  non-relativistic particles give negligible contributions.
Moreover,
the thermal vacuum of the hot plasma filling the center of galaxy clusters, which  has temperatures of order of $(10 \div 100) \times 10^{6}K $,
has an energy density of $(10^{-48} - 10^{-47})GeV^4$
 and a state equation $w = 0.01$. Such values of $\rho$ and $w$ are in agreement with the ones of the dark matter.

The thermal states can describe also the  Unruh  and of the Hawking effects;  however,
both of the  phenomena do not contribute to the energy of the universe, since the temperatures are very low \cite{CapolupoCMB:2016}.

\section{V. Vacuum contribution of fields in curved background}

Fields in curved background are also characterized by condensed vacuum and Bogoliubov transformations \cite{Birrell}.
For such fields, the energy density and pressure depend on the particular metric considered.
Here one  considers the spatially flat Friedmann Robertson-Walker metric
$
d s^{2} = d t^{2} - a^{2}(t) d {\bf x}^{2} = a^{2}(\eta) (d \eta^{2}- d {\bf x}^{2})\,,
$
where $a$ is the scale factor, $t$ is the comoving time, $\eta$ is the conformal time, $\eta(t) = \int_{t_{0}}^{t} \frac{d t }{a(t)}$, with $t_0$ arbitrary constant.

The energy density and pressure are expressed as \cite{Parker1,Parker2}
\bea\non
\rho_{curv} &=& \frac{2 \pi}{a^2}\int_{0}^{K} dk k^{2} \lf(|\phi_{k}^{\prime}|^{2} + k^{2} |\phi_{k}|^{2} + m^{2} |\phi_{k}|^{2} \ri),
\\
\\\non
p_{curv} &=& \frac{2 \pi}{a^2}\int_{0}^{K} dk k^{2} \lf(|\phi_{k}^{\prime}|^{2} - \frac{k^{2}}{3} |\phi_{k}|^{2} - m^{2} |\phi_{k}|^{2} \ri).
\\
\eea
where $K$ is the cut-off on the momenta, $\phi_{k}$ are mode functions and  $\phi_{k}^{\prime}$  denotes the derivative of $\phi_{k}$ with respect to the conformal time  $\eta$.
Assuming  at late time the cutoff on the momenta much smaller than the comoving mass of the field, $K \ll m a$ and setting $m \gg H$,  for an arbitrary Robertson-Walker metric in infrared regime, one has \cite{maroto}
\bea\non
\rho_{curv} &=& \frac{1}{8 \pi^2} \int_{0}^{K} dk k^{2} \lf(\frac{2 m}{a^{3}} + \frac{9 H^{2}}{4 m a^{3}} + \frac{k^2}{m a^{5}} \ri),
\\
\\\non
p_{curv} &=& \frac{1}{8 \pi^2}\int_{0}^{K} dk k^{2} \lf( \frac{9 H^{2}}{4 m a^{3}}  - \frac{k^2}{3 m a^{5}} \ri).
\\
\eea
The state equation is  $w_{curv} \simeq 0$, which coincides with the one of the dark matter.
Numerical  values compatible with the ones of dark matter  are found when
$\frac{m K^{3}}{ a^{3} } \sim 10^{-45}GeV^{4}$.

\section{VI. Vacuum contributions of Particle mixing}

 The  particle  mixing concerns  neutrinos and quarks in fermion sector,  axions, kaons, $B^0$, $D^0$, and $\eta-\eta^\prime$ systems, in boson sector. In the case of mixing between two fields, it is expressed as
  \bea\label{mixing}\non
 \varphi_{1}(\theta,x) &=&     \varphi_{1}(x) \cos(\theta) + \varphi_{2}(x) \sin(\theta)\,,
\\
 \varphi_{2}(\theta,x) &=&  - \varphi_{1}(x) \sin(\theta) + \varphi_{2}(x) \cos(\theta)\,,
 \eea
 where, $\theta$ is the mixing angle, $ \varphi_{i}(\theta,x)$ are the mixed fields and $ \varphi_{i}(x)$ are the free fields, with $i =1,2$.

The mixing transformations (\ref{mixing}) and the mixed
annihilation operators can be expressed by means of the generator $J (\theta, t)$ as $\varphi_{i}(\theta,x) \equiv J^{-1}(\theta, t) \varphi_{i}(x) J (\theta, t)$,   and $\chi^{r}_{{\bf k},i}(\theta, t)
\equiv J^{-1}(\theta, t)\;\chi^{r}_{{\bf k},i}(t)\;J (\theta, t)$, respectively, with
$\chi^{r}_{{\bf k},i} = a_{{\bf k},i}, \alpha^{r}_{{\bf k},i}$, for bosons and fermion, respectively, and $i =1,2$ \cite{Blasone:2002jv,Blasone:2001du}.

The  physical vacuum where particle oscillations appears  is $|0(\theta, t)\ran \,\equiv\,J^{-1}(\theta, t)\;|0\ran_{1,2} $, where
$|0\ran_{1,2}$ is the vacuum annihilated by $\chi^{r}_{{\bf k},i} $. One has
\bea \label{con}   \langle 0(\theta, t)| \chi_{{\bf k},i}^{r \dag} \chi^r_{{\bf
k},i} |0(\theta, t)\rangle  = \sin^{2}\te ~ |\Upsilon^{\lambda}_{{\bf
k}}|^{2},
\eea
where $\lambda = B,F$, $i=1,2$ and the reference frame ${\bf k}=(0,0,|{\bf k}|)$ has
been adopted. The Bogoliubov
coefficient entering the mixing transformation $\Upsilon^{\lambda}_{{\bf
k}}$ assumes the following form for boson and fermion
\bea\label{Bogoliubov}
| \Upsilon^{B}_{{\bf
k}}|  &=&   \frac{1}{2} \lf( \sqrt{\frac{\Omega_{k,1}}{\Omega_{k,2}}} -
\sqrt{\frac{\Omega_{k,2}}{\Omega_{k,1}}} \ri)\,,
\\
|\Upsilon^{F}_{{\bf
k}}|  &=&  \frac{ (\Omega_{k,1}+m_{1}) - (\Omega_{k,2}+m_{2})}{2
\sqrt{\Omega_{k,1}\Omega_{k,2}(\Omega_{k,1}+m_{1})(\Omega_{k,2}+m_{2})}}\, |{\bf k}| \,,
\eea
respectively,
where $| \Sigma^B_{{\bf k}}|^{2}  - | \Upsilon^B_{{\bf k}}|^{2}  = 1 $ and $| \Sigma^F_{{\bf k}}|^{2}  + | \Upsilon^F_{{\bf k}}|^{2}  = 1 $,
($\Sigma^\lambda_{{\bf k}}$ are the other coefficients entering in the transformations), $\Omega_{k,i}$ are the energies of the free fields,  $i=1,2$.

- \emph{Boson mixing} -
The energy density and pressure of the vacuum condensate induced by mixed bosons are

\begin{widetext}
 \bea \label{T00BosMix}
\rho^{B}_{mix} & = &    \frac{1}{2} \langle 0| : \sum_i \Big[\pi^{2}_{i}(-\theta,x) + \lf[\vec{\nabla} \phi_{i}(-\theta,x)  \ri]^{2}
+ m^{2}_{i} \phi^{2}_{i}(-\theta,x) \Big]: |0 \rangle\,;
\\\non
\\ \label{TjjBosMix}
p^{B}_{mix}
& = &  \langle 0| :\sum_i  \Big(\lf[ \partial_{j} \phi_{i}(-\theta,x) \ri]^{2}+ \frac{1}{2}\Big[\pi_{i}^{2}(-\theta,x)
 -
  \lf[\vec{\nabla} \phi_{i}(-\theta,x)  \ri]^{2}
 - m_{i}^{2}
\phi_{i}^{2}(-\theta,x)  \Big] \Big): |0 \rangle\,,
 \eea
 \end{widetext}
respectively.
One can see that   the kinetic and gradient terms of the mixed vacuum are equal to zero \cite{CapolupoDark}
  \bea \non
  && \langle 0| : \sum_i  \pi^{2}_{i}(-\theta,x):|0 \rangle\, =  \langle 0| : \sum_i   \lf[\vec{\nabla} \phi_{i}(-\theta,x)  \ri]^{2}
 : |0 \rangle\,
 \\\non
 && = \langle 0| :\sum_i  \lf[ \partial_{j} \phi_{i}(-\theta,x) \ri]^{2}:|0 \rangle\, = 0\,.
  \eea
  Then, Eqs.(\ref{T00BosMix}) and (\ref{TjjBosMix}) reduce to
   \bea \label{T00BosMixF}
\rho^{B}_{mix} & = &    \langle 0| :\sum_{i}
  m^{2}_{i}
\phi_{i}^{2}(-\theta,x) : |0 \rangle\,,
\\ \label{TjjBosMixF}
p^{B}_{mix} & = & -    \langle 0| :\sum_{i}
  m^{2}_{i}
\phi_{i}^{2}(-\theta,x) : |0 \rangle\,,
 \eea
    and the state equation  coincides with the ones of the cosmological constant, $w^{B}_{mix} = - 1$.
 By setting $\Delta m^{2}= |m_{2}^{2}-m_{1}^{2}|$, one has
 \bea\label{integral}
 \rho^{B}_{mix} = \frac{\Delta m^{2}   \sin ^{2}\theta}{8 \pi^{2}} \int_{0}^{K} dk k^{2} \lf(\frac{1}{\omega_{k,1}}- \frac{1}{\omega_{k,2}}\ri),
 \eea
where $K$ is the cut-off on the momenta.

- In the case of axion-photon mixing, for magnetic field strength $B \in [10^{6} - 10^{17}] G$, axion mass $m_{a} \simeq 2 \times 10^{-2}eV$,   $\sin^{2}_{a}\theta \sim 10^{-2}$ and  a Planck scale cut-off, $K\sim 10^{19} GeV$, one obtain a value of the energy density $\rho^{axion}_{mix} = 2.3 \times 10^{-47}GeV^{4}$,
 which is of the same order of  the estimated upper bound on the dark energy.

-  In the case of superpartners of the neutrinos, considering masses  $m_1 = 10^{-3} eV$ and $m_2 = 9 \times 10^{-3} eV$, such that $\Delta m^{2} = 8 \times 10^{-5} eV^{2}$ and assuming
$\sin^{2} \theta = 0.3$, one obtains, $\rho^{B}_{mix}  = 7 \times 10^{-47}GeV^{4}$ for a cut-off on the momenta  $K =10  eV$.
  Smaller values of the mixing angle lead to values of  $\rho^{B}_{mix}$ which are compatible with the estimated value of the dark energy also in the case in which the cut-off is $K = 10^{19} GeV$, indeed $\rho^{B}_{mix}$ depends linearly by  $\sin^{2}\theta$ \cite{CapolupoDark}.

- \emph{Fermion mixing} - The energy density and pressure are
\begin{widetext}
\bea\label{T00FerMix}
\rho_{mix}^{F} & = & -   \langle 0| : \sum_i \Big[\psi_{i} ^{\dag}(-\theta,x)\gamma_{0} \gamma^{j} \partial_{j} \psi_{i}(-\theta,x)
+ m
\psi_{i}^{\dag}(-\theta,x) \gamma_{0}  \psi_{i}(-\theta,x)\Big]: |0 \rangle\,;
\\\label{TjjFerMix}
p_{mix}^{F}  & = &  i    \langle 0| :\sum_i \Big[\psi_{i}^{\dag}(-\theta,x)\gamma_{0} \gamma_{j} \partial_{j} \psi_{i}(-\theta,x) \Big]: | 0 \rangle \,,
 \eea
 \end{widetext}
 where $\psi_{i}(-\theta,x)$ are the flavor neutrino fields or the quark fields.
Being
\bea\non
 && \langle 0| : \sum_i \bar{\psi}_{i}(-\theta,x)\gamma^{j} \partial_{j} \psi_{i}(-\theta,x): | 0 \rangle \ =
 \\\non
 && \langle 0| : \sum_i \Big[\psi_{i}^{\dag}(-\theta,x)\gamma_{0} \gamma_{j} \partial_{j} \psi_{i}(-\theta,x)\Big]: | 0 \rangle \ = 0\,,
\eea
one has
\bea\label{T00FerMixF}
\rho_{mix}^{F}  & = & -   \langle 0| : \sum_i \Big[m_i
\psi_{i}^{\dag}(-\theta,x) \gamma_{0}  \psi_{i}(-\theta,x)\Big]: |0 \rangle,
\\\label{TjjFerMixF}
p_{mix}^{F}  & = &  0 \,.
 \eea
The  state equation  is then $w^{F}_{mix} =0$, which is the one of the dark matter.
 The   the energy density  is
\bea\label{ener-Fer}\non
\rho^{F}_{mix} = \frac{\Delta m \sin ^{2}\theta}{2 \pi^{2}} \int_{0}^{K} dk k^{2}
\lf( \frac{m_2 }{ \omega_{k,2}} - \frac{m_1}{ \omega_{k,1}}\ri)\Big]\,.
\eea
By considering  masses of order of $10^{-3}eV$, such that $\Delta m^{2} \simeq 8 \times 10^{-5}eV^2$ and a cut-off on the momenta  $K = m_1 + m_2$, one obtains  $\rho^{F}_{mix} = 4 \times 10^{-47 } GeV^4$, which is in agreement with the estimated upper bound of the dark matter.
For $K$ of order of the Plank scale one has  $\rho^{F}_{mix} \sim  \times 10^{-46 } GeV^4$.
Notice that the quark confinement inside the hadrons should inhibit the gravitational interaction of the quark vacuum condensate. Thus the quark condensate should not affect the dark matter of the universe.

\section{VII. SUSY breaking and vacuum condensate}

We show that  vacuum condensate provides a new mechanism of spontaneous SUSY breaking.
We start by a situation in which SUSY is preserved at the  lagrangian level and study the effects of vacuum condensation.
 We consider a Bogoliubov transformation acting
  simultaneously and with the same parameters on the bosonic and on the fermionic degrees of freedom in order not to break SUSY explicitly.
Since,    in any field theory which has manifest supersymmetry at the lagrangian level, a nonzero vacuum energy implies the spontaneous SUSY breaking \cite{Witten:1981nf},  then the vacuum condensate (which is characterized by nontrivial energy) breaks  SUSY  spontaneously.

The effects of a Bogoliubov transformation are analyzed in the Wess--Zumino model described by the Lagrangian \cite{Wess:1973kz}
\bea \label{WS}\non
\mathcal{L} &=& \frac{i}{2} \bar{\psi}\gamma_{\mu}\partial^{\mu}\psi + \frac{1}{2}\partial_{\mu}S\partial^{\mu}S + \frac{1}{2}\partial_{\mu}P\partial^{\mu}P
\\ &-& \frac{m}{2}  \bar{\psi}\psi - \frac{m^2}{2} (S^2 + P^2),
\eea
where $\psi$ is a Majorana spinor field, $S$ is a scalar field and $P$ is a pseudoscalar field. This Lagrangian is invariant under supersymmetry transformations \cite{Wess:1973kz}.

The fields are quantized by expanding them in modes:
\bea
\psi(x)&=& \sum_{r=1}^2\int \frac {d^3\mathbf{k}}{(2\pi)^{\frac{3}{2}}}\,\, e^{i \mathbf{k}\mathbf{x}}\lf[u^r_{\mathbf{k}}\alpha^r_{\mathbf{k}}(t)
+ v^r_{-\mathbf{k}}\alpha^{r \dagger}_{\mathbf{-k}}(t)\ri],\\
S(x)&=& \int \frac {d^3\mathbf{k}}{(2\pi)^{\frac{3}{2}}}\,\, \frac{1}{\sqrt{2 \omega_{k}}} \,\, e^{i \mathbf{k}\mathbf{x}}\lf[b_{\mathbf{k}}(t)
+ b^{\dagger}_{\mathbf{-k}}(t)\ri],\\
P(x)&=& \int \frac {d^3\mathbf{k}}{(2\pi)^{\frac{3}{2}}}\,\, \frac{1}{\sqrt{2 \omega_{k}}} \,\, e^{i \mathbf{k}\mathbf{x}}\lf[c_{\mathbf{k}}(t)
+ c^{\dagger}_{\mathbf{-k}}(t)\ri].
\eea
where $\alpha^r_{\mathbf{k}}(t) = \alpha^r_{\mathbf{k}} e^{-i\omega_{k}t}$, $b_{\mathbf{k}}(t) = b_{\mathbf{k}} e^{-i\omega_{k}t}$, $c_{\mathbf{k}}(t) = c_{\mathbf{k}} e^{-i\omega_{k}t}$ and $\omega_{\mathbf{k}}=\sqrt{k^2 + m^2}$, and since $\psi$ is a Majorana spinor, one has $v^r_{\mathbf{k}}=\gamma_0 C (u^r_{\mathbf{k}})^*$ and $u^r_{\mathbf{k}}=\gamma_0 C (v^r_{\mathbf{k}})^*$.

The vacuum annihilated by $\alpha^r_{\mathbf{k}}$, $b_{\mathbf{k}}$ and $c_{\mathbf{k}}$ is defined as $|0\rangle=|0\rangle^{\psi}\otimes|0\rangle^S\otimes|0\rangle^P$.

We perform simultaneous Bogoliubov transformations  on   fermion and   boson  annihilators,
\bea\label{Bog1}
 {\alpha}^r_{\mathbf{k}}(\xi, t) &=& U^{\psi}_{\mathbf{k}} \, \alpha^r_{\mathbf{k}}(t) + V^{\psi}_{-\mathbf{k}} \, \alpha^{r\dagger}_{-\mathbf{k}}( t)\,, \non
\\
 {\alpha}^{r\dagger}_{-\mathbf{k}}(\xi,  t) &=& U^{\psi *}_{-\mathbf{k}} \, \alpha^{r\dagger}_{-\mathbf{k}}(t) + V^{\psi *}_{ \mathbf{k}} \, \alpha^{r}_{\mathbf{k}}(t)\,,
\\[2mm]\non
 {b}_{\mathbf{k}}(\eta,  t) &=& U^{S}_{\mathbf{k}} \, b_{\mathbf{k}}(t) - V^{S}_{-\mathbf{k}}  \,b^{\dagger}_{-\mathbf{k}}(t)\,,
\\\label{Bog2}
 {b^{\dag}}_{-\mathbf{k}}(\eta,  t) &=& U^{S *}_{-\mathbf{k}} \, b^{\dag}_{-\mathbf{k}}(t) - V^{S *}_{\mathbf{k}}  \,b_{\mathbf{k}}(t)\,,
\\[2mm]\non
 {c}_{\mathbf{k}}(\eta,  t) &=& U^{P}_{\mathbf{k}} \, c_{\mathbf{k}}(t) - V^{P}_{-\mathbf{k}}  \,c^{\dagger}_{-\mathbf{k}}(t)\,,
\\\label{Bog3}
 {c^{\dag}}_{-\mathbf{k}}(\eta,  t) &=& U^{P *}_{-\mathbf{k}} \, c^{\dag}_{-\mathbf{k}}(t) - V^{P *}_{\mathbf{k}}  \,c_{\mathbf{k}}(t)\,.
\eea
One has,  $U^{S}_{\mathbf{k}} = U^{P}_{\mathbf{k}} $ and $V^{S}_{\mathbf{k}} =V^{P}_{\mathbf{k}} $.
We denote such quantities as $U^{B}_{\mathbf{k}} $ and $V^{B}_{\mathbf{k}} $, respectively.

Eqs.(\ref{Bog1})--(\ref{Bog3}) can be written as
\bea
{\chi}^r_{\mathbf{k}}( \xi,  t) &=& J^{-1} (\xi,\eta,  t)\,\chi^r_{\mathbf{k}}(t) J(\xi,\eta,  t)\,,
\eea
with $\chi_{\mathbf{k}} = \alpha_{\mathbf{k}}, b_{\mathbf{k}},  c_{\mathbf{k}} $  and
$
J(\xi,\eta,   t) = J_{\psi}(\xi,  t) J_{S}(\eta,  t) J_{P}(\eta,  t)\,,
$
where
\bea\non
J_{\psi} &=& \exp \lf[\frac{1}{2}\int d^{3} \mathbf{k} \, \xi_{\mathbf{k}}(\zeta)\lf(\alpha^r_{\mathbf{k}}(t) \alpha^r_{-\mathbf{k}}(t)  - \alpha^{r \dagger}_{-\mathbf{k}}(t) \alpha_{ \mathbf{k}}^{r \dagger}(t)  \ri)\ri],
\\\non
J_{S} &=& \exp \lf[-i\int d^{3} \mathbf{k}\, \eta_{\mathbf{k}}(\zeta)\lf(b_{\mathbf{k}}(t) b_{-\mathbf{k}}(t)  - b_{-\mathbf{k}}^{\dagger}(t) b_{ \mathbf{k}}^{\dagger}(t)  \ri)\ri],
\\\non
J_{P} &=& \exp \lf[-i\int d^{3} \mathbf{k}\, \eta_{\mathbf{k}}(\zeta)\lf(c_{\mathbf{k}}(t) c_{-\mathbf{k}}(t)  - c_{-\mathbf{k}}^{\dagger}(t) c_{ \mathbf{k}}^{\dagger}(t)  \ri)\ri].
\\
\eea

The transformed vacuum  is
$| {0}( \xi,\eta,   t)\rangle=| {0}( \xi,   t)\rangle_{\psi}\otimes | {0}(\eta,  t)\rangle_{S}\otimes | {0}( \eta,  t)\rangle_{P}$, where
$
| {0}( \xi,  t)\rangle_{\psi} = J^{-1}_{\psi}(\xi,  t)|0\rangle_{\psi} $,
$ | {0}(\eta,   t)\rangle_{S} = J^{-1}_{S}(\eta,  t)|0\rangle_{S}$,
$  | {0}(\eta,    t)\rangle_{P} = J^{-1}_{P}(\eta,  t)|0\rangle_{P}
$, respectively,
and then
\bea
| {0}( \xi,\eta,  t)\rangle = J^{-1}(\xi,\eta,  t)|0\rangle\,.
\eea
In a supersymmetric context,
$| {0}( \xi,\eta,  t)\rangle $  is the physical vacuum for the systems listed above.
 The nonzero energy of $| {0}( \xi,  t)\rangle_{\psi} $, $ | {0}(\eta,   t)\rangle_{S}$  and $ | {0}(\eta,    t)\rangle_{P} $, (see above)
leads to an energy densities different from zero for $| {0}( \xi,\eta,   t)\rangle$.
Indeed, considering the free Hamiltonian $H$ corresponding to the Lagrangian in Eq.(\ref{WS}), $H= H_{\psi} +  H_B $ where $H_B = H_S + H_P$,
the expectation values of the fermion  and boson Hamiltonians on $|{0}( \xi,\eta,  t)\rangle$  are
\bea\non\label{Hpsiasp}
&&\langle{0}( \xi,\eta,  t)| H_{\psi} |{0}( \xi,\eta,  t)\rangle =
- \int \; d^3\mathbf{k}\; \omega_{\mathbf{k}} \,(1- 2 |V^{\psi}_{\mathbf{k}}|^2)\,,
\eea
and
\bea\label{HBasp}
\langle{0}( \xi,\eta,  t)| H_B |{0}( \xi,\eta,  t)\rangle = \int\; d^3\mathbf{k}\; \omega_{\mathbf{k}} (1 + 2 |V^{B}_{\textbf{k}}|^2)\,,
\eea
respectively. Then   we have the  final result
\bea\label{Ht}
\langle\tilde{0}(  t)| H |\tilde{0}(  t)\rangle = 2 \int\; d^3\mathbf{k}\; \omega_{\mathbf{k}} (|V^{\psi}_{\textbf{k}}|^2 + |V^{B}_{\textbf{k}}|^2)\,,
\eea
which is different from zero and positive.

Notice that, Eq.(\ref{Ht}), holds for disparate physical phenomena. As remarked above, the explicit form of the Bogoliubov coefficients
$V^{\psi}_{\textbf{k}}$ and $V^{B}_{\textbf{k}}$
specifies the particular system.

Laser cooling experiments could allow to test the mechanism of SUSY breaking here presented.   Indeed,   the Wess-Zumino model in $2+1$ dimensions can be obtained by a mixture of cold atoms-molecules trapped in two dimensional optical lattices \cite{Yue Yang}. In this case SUSY is preserved at zero temperature and is broken at $T\neq 0$.
Then, a signature of SUSY breaking in such a system can be probed   by the detection of the constant background noise due to the nonzero energy of the thermal vacuum, given by (\ref{Ht}). In the thermal case,   $V^{B}_{\textbf{k}} = \sqrt{\frac{1}{{e^{\beta \omega_{\textbf{k}}}}-1}}$ and $V^{\psi}_{\textbf{k}} = \sqrt{\frac{1}{{e^{\beta \omega_{\textbf{k}}}}+1}}$, where $\beta$ is the inverse temperature (in units such that $k_B=1$).
Considering a two-dimensional optical lattice and a tuning of the parameters such that the effective mass of the emerging fields is zero, the vacuum energy density, due to Eq.(\ref{Ht}), and representing the background noise,   is given by
$<H>\,=\,14 \, \pi\,\zeta(3)\, T^{3} $.

\section{VIII. Conclusions}

The vacuum condensates characterizing many systems can contribute to the dark matter and to the dark energy.
Dark matter contributes derive by thermal vacuum of  intercluster medium, by the vacuum of   fields in curved space  and by the neutrino flavor vacuum. Dark energy contribute are given by axion-photon mixing.
We have also shown that, in a supersymmetric field theory, vacuum condensates may leads to  spontaneous SUSY breaking.

\section{Conflict of Interests}

The authors declare that there is no conflict of interests
regarding the publication of this paper.

\section{Acknowledgements}
Partial financial support from MIUR is acknowledged.

\end{document}